\documentclass{ws-fnl}

\usepackage{xcolor}
\usepackage[verbose]{hyperref}
\hypersetup{colorlinks=false,allbordercolors=blue,pdfborderstyle={/S/U/W 1}}

\begin{document}

\markboth{S.~A.~Flanery and C.~Chamon}{Noise-Based Authentication: Is It Secure?}

\catchline{}{}{}{}{}

\title{Noise-Based Authentication: Is It Secure?}

\author{Sarah A. Flanery}

\address{Department of Electrical and Computer Engineering, Texas A\&M University\\
College Station, Texas 77843,
USA\\
sflanery@tamu.edu}

\author{Christiana Chamon\footnote{Corresponding author}}

\address{Department of Electrical and Computer Engineering, Virginia Tech\\
Blacksburg, Virginia 24061, USA\\
ccgarcia@vt.edu}

\maketitle

\begin{history}
\received{(Day Month Year)}
\revised{(Day Month Year)}
\accepted{(Day Month Year)}
\published{(Day Month Year)}
\comby{}
\end{history}

\begin{abstract}
This paper introduces a three-point biometric authentication system for a blockchain-based decentralized identity network. We use existing biometric authentication systems to demonstrate the unique noise fingerprints that belong to each individual human and the respective information leak from the biological characteristics. We then propose the concept of using unique thermal noise amplitudes generated by each user and explore the open questions regarding the robustness of unconditionally secure authentication.
\end{abstract}


\section{Introduction}

\subsection{Motivation: Decentralized Identity}

The internet that we use today is based on Web 2.0, where each user’s identity is tied to a centralized source. Users do not have ownership of their data, for in the event the centralized source were to vanish, the data would no longer exist. Web 3.0 eliminates this problem entirely by introducing Decentralized Identity protocols, i.e. an authentication scheme that does not require a central authority for identification [1-5].

Decentralized Identity architecture typically consists of an Ethereum blockchain for the purpose of appending blocks tied to each user’s specific decentralized identifier (DID). A byproduct of this public ledger is that DID users have complete ownership of their data because centralized figures cannot regulate them. As a result of not having to rely on third-party sources, users not only have all their data in a singular place, but also as a method to decrease potential corruption.

\begin{figure}
    \centering
    \includegraphics{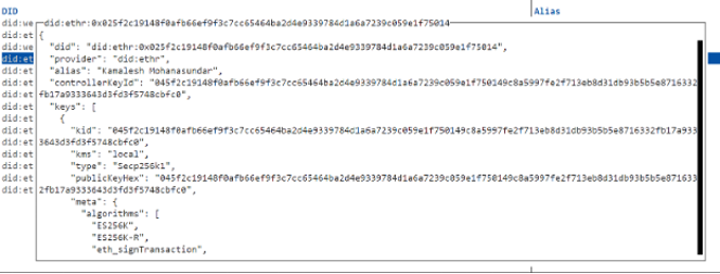}
    \caption{Example DID document for user “Kamalesh Mohanasundar” consisting of the domain of the ledger (“did:ethr”), specifically-assigned DID (“did”), and respective public key (“publicKeyHex”).}
    \label{didexample}
\end{figure}

Each Ethereum DID is generated using asymmetric cryptography, consisting of a private key pair and a hash [1-3]. The DID document containing the identity of the user is then verified, either externally or by the user. In the latter case, biometric login such as fingerprinting or face ID is often used, as it cannot be easily replicated by outside persons. After verification, the DID document, such as the one displayed in Figure~\ref{didexample}, is appended to the blockchain.

Recently, Flanery et al. published work on a zero-trust ecosystem that uses decentralized identity (DID) protocols for ownership of learning credentials [2, 3]. Each user has their own unique decentralized identifier, and all of their knowledge badges are stored on an Ethereum blockchain. While the system is secure internally, the users authenticate themselves with a username and password, which poses inherent security risks such as password reuse or weak passwords, phishing attacks, credential-stuffing, database breaches, and brute-force attacks [1-16].

\subsection{The Importance of User Authentication}

In a peer-to-peer network, it is important to establish that the user is who they say they are in order to maintain the confidentiality, integrity, and accessibility of login systems. This is to prevent threats that can lead to unauthorized reading and writing of the user’s data, as well as denial-of-service (DoS) attacks against users. Authentication is essential due to the amount of data exchanged within peer-to-peer networks on a regular basis, as well as the recent year-to-year increase in the number of data breaches, typically as a result of carelessness [4, 7-9]. The security of authentication systems involves ensuring that all physical layers are robust against adversarial attacks, such that users are not unknowingly logging in and exchanging information.

Common methods of login include multi-factor authentication, tokens, certificates, single sign-on, behavioral, devices, or a combination thereof. One such method is biometric authentication, e.g. fingerprint, retina, and facial recognition; they provide a higher level of security compared to traditional passwords or PINs, as physical traits are much harder to duplicate compared to passwords or ID cards. Additionally, the ease of use and speed of biometric systems enhance the overall user experience, as users appreciate the seamless and quick authentication process in mobile and high-security environments [12-25].

Multi-layer biometric authentication enhances security by combining multiple biometric traits to verify a user's identity; it becomes much harder for attackers to spoof multiple biometric traits simultaneously, and multi-layer systems cross-verify multiple traits, reducing the chances of false acceptances (incorrectly granting access) and false rejections (incorrectly denying access). Combining multiple biometrics can streamline the authentication process; if one method fails (e.g. a fingerprint is not recognized due to a cut), another method (e.g. facial recognition) can provide a backup. Most importantly, relying on multiple biometrics reduces the dependency on a single authentication factor, and this diversification enhances overall system resilience against attacks or failures in any one biometric method [24, 25].

\subsection{Noise Fingerprinting}

Noise fingerprinting is a method of authentication that leverages the unique noise characteristics inherent in various electronic devices or transmission channels. These noise characteristics, often referred to as “noise fingerprints,” are random but reproducible patterns that can be used to uniquely identify a device or a communication session. This technique is gaining popularity in the field of cybersecurity for its ability to provide a high level of security and difficulty in being spoofed [18-21].

Examples of noise fingerprinting include electronic noise patterns in circuits, variations in sound profiles, and network traffic patterns. The inherent uniqueness, non-intrusiveness, and robustness render it useful for device authentication, secure communications, and access controls. Recent efforts have been made to improve the accuracy and robustness of noise-based authentication systems to enhance feature extraction and matching processes [1, 18-21].

\subsection{Biometric Authentication}

Each individual's noise pattern in fingerprint readers is unique, and facial recognition is also unique to each person (except for identical twins). Similarly, unique eye patterns generate distinct noise amplitudes through eye movement. By analyzing noise patterns from these three biometric points, user authentication can be validated only when all three unique patterns are verified.

In the present paper, we propose a three-layer authentication scheme using noise radiated by each user. We choose three layers to account for both added redundancy and time elapsed on authentication. We hypothesize that each user produces a unique noise pattern, and we can match the noise fingerprint to each individual for more robust authentication.

The rest of this paper is organized as follows. Section 2 describes the methodology for extracting noise fingerprints from existing biometric systems, Section 3 summarizes the results obtained from Section 2, and Section 4 raises the open question of unconditionally secure statistical physical authentication.

\section{Methodology}

\begin{figure}
    \centering
    \includegraphics[scale=0.75]{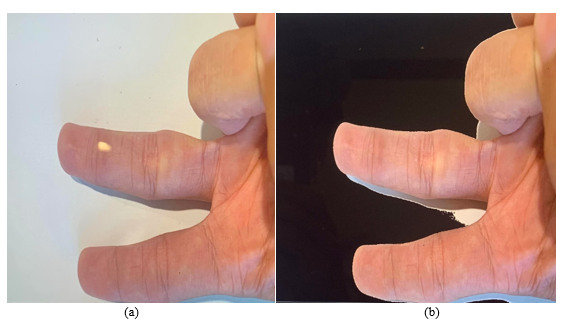}
    \caption{An example of (a) a user’s fingers (facing up) against a white background and (b) the resulting image after applying a mask to the skin area.}
    \label{fingerprint_example}
\end{figure}

The Chamon Authentication System (CAS) consists of three layers of biometric authentication: fingerprint-reading, face-recognizing, and eye-tracking. We used a webcam to capture images of user fingertips against a white background. The image is represented by individual pixels, referred to as "frames," each with unique R, G, and B components. We summed these components for each frame and assigned the resulting RGB value using the Cascade Classifier mapping library [26]. To isolate the fingerprint and ensure that only the fingerprint’s RGB values are captured, we apply a mask to the skin area, as demonstrated by the example in Figure~\ref{fingerprint_example}.

\begin{figure}
    \centering
    \includegraphics[scale=0.75]{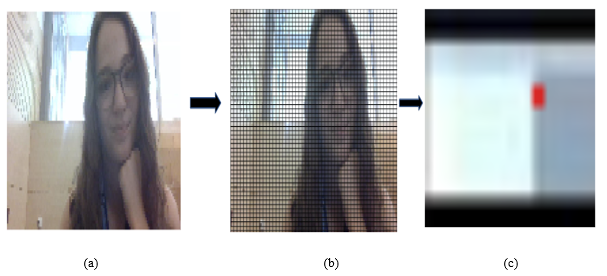}
    \caption{An example of (a) a photograph of a user’s face, (b) division of a photograph into frames, and (c) a single frame being assigned an RGB value.}
    \label{Facecam_example}
\end{figure}

Similarly to the fingerprint method, we used a webcam to capture images of user faces and assigned an RGB summation value to each frame, as illustrated by the example in Figure~\ref{Facecam_example}. We then use Face Cascade [26], to isolate the user’s face from the background.

\begin{figure}
    \centering
    \includegraphics[scale=0.75]{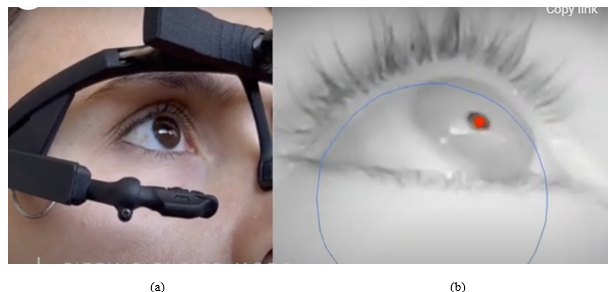}
    \caption{An example of the Pupil Core eye-tracker in use. The eye-tracker rests over the user’s eyes (a) and tracks the changes in the user’s eye movement from the center of the pupil (b).}
    \label{Eye_Tracker_example}
\end{figure}

For the eye-tracking, we place the Pupil Core eye-tracker on the user, activate light stimuli in front of their pupil, and gather the involuntary pupil movements away from the resting pupil position, as shown in Figure~\ref{Eye_Tracker_example}. We do this for the purpose of measuring the noise emitted by a user's eyes.

\section{Summary of Results}
\subsection{Fingerprint Demonstration}

\begin{figure}
    \centering
    \includegraphics{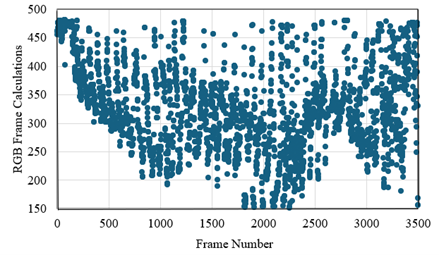}
    \caption{Scatter plot of the RGB frame values from the fingerprint in Figure~\ref{fingerprint_example}, with respect to the individual frames. These values are unique to the user.}
    \label{Fingerprint_Demonstration}
\end{figure}

Figure~\ref{Fingerprint_Demonstration} shows the RGB values for the fingerprint in Figure~\ref{fingerprint_example} with respect to the individual frames. The RGB values, represented by the blue dots, are unique to the user. 

\begin{figure}
    \centering
    \includegraphics{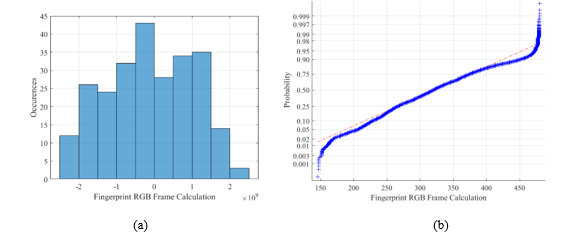}
    \caption{Histogram (a) and normal-probability plot (b) of the data in Figure~\ref{Fingerprint_Demonstration}. The values appear to have a distribution similar to normal, besides the deviating tail ends.}
    \label{Fingerprint_Data}
\end{figure}

The histogram (a) and probability plot (b) of the RGB value with respect to the frame number is shown in Figure~\ref{Fingerprint_Data}. While the data follows a bell-like distribution, the tails appear to deviate from Gaussianity, as they provide the most information about the user.

\subsection{Face ID Demonstration}

\begin{figure}
    \centering
    \includegraphics{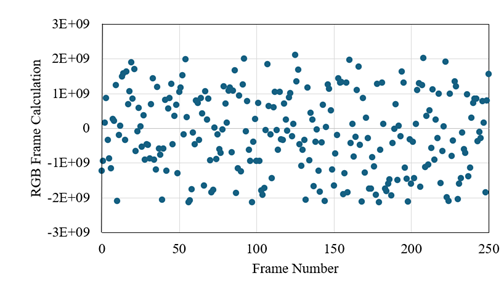}
    \caption{Scatter plot of the RGB frame values from the facial image in Figure~\ref{Facecam_example}, with respect to the individual frames. These values are unique to the user.}
    \label{Facecam_Demo}
\end{figure}

\begin{figure}
    \centering
    \includegraphics{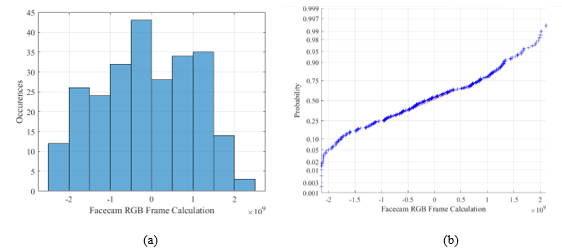}
    \caption{Histogram (a) and normal-probability plot (b) of the data in Figure~\ref{Facecam_Demo}. The values appear to have a distribution like normal, besides the deviating tail ends.}
    \label{Facecam_Data}
\end{figure}

Figure~\ref{Facecam_Demo} shows the unique RGB values of a user's face, with respect to the frame number. Figure~\ref{Facecam_Data} shows the histogram (a) and normal-probability plot (b) of the data in Figure~\ref{Facecam_Demo}. The values appear to have a distribution similar to normal, besides the deviating tail ends. Similarly to the fingerprint demonstration (see Section 3.1), the values appear to have a distribution like normal, besides the deviating tail ends.

\subsection{Eye-Tracker Demonstration}

\begin{figure}
    \centering
    \includegraphics[scale=0.75]{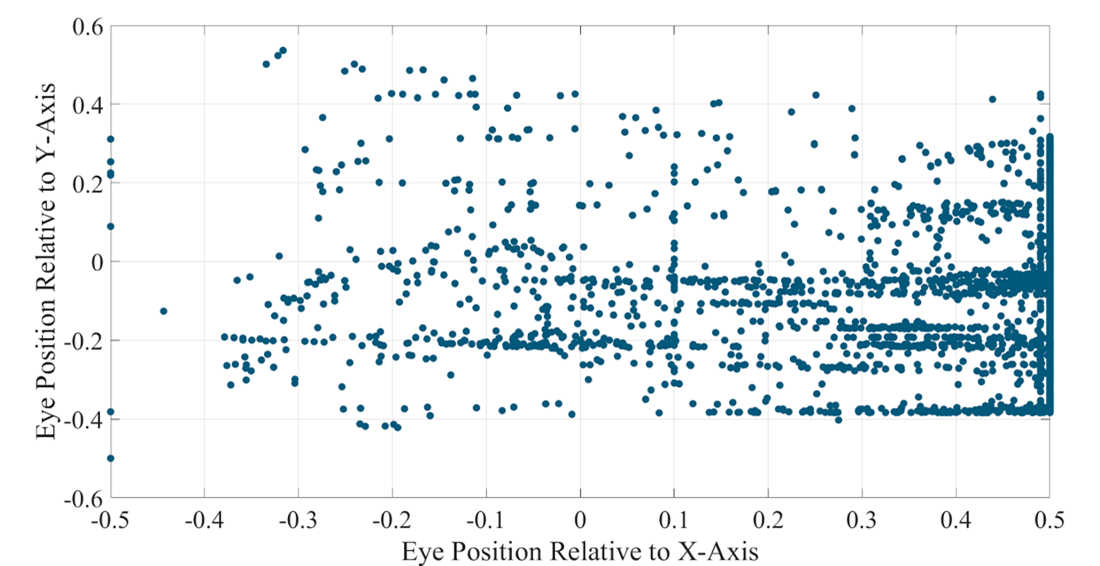}
    \caption{Scatter plot of the changes in a user’s eye position. Note that the overwhelming majority of the data points are grouped together in what appears to be a vertical line.}
    \label{Eyetracker_Demo}
\end{figure}

Figure~\ref{Eyetracker_Demo} shows the change in eye positions after the user was shown a light stimulus (see Section 2). The x position of the eye is relatively stationary, as there is little movement to be observed. However, the pupil reacted to changes in the y position [27].

\begin{figure}
    \centering
    \includegraphics{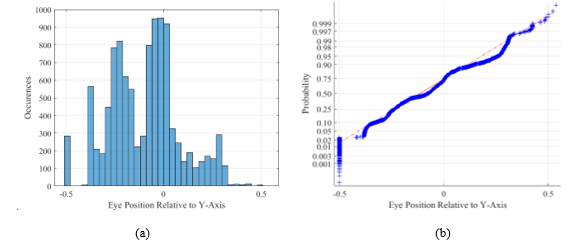}
    \caption{Histogram (a) and normal-probability plot (b) of the data in Figure 8. The values appear to have a distribution like normal, besides the deviating tail ends.}
    \label{Eyetracker_Data}
\end{figure}

The histogram (a) and probability plot (b) of the RGB value with respect to the frame number is shown in Figure~\ref{Eyetracker_Data}. While the data follows a bell-like distribution, the tails appear to deviate from Gaussianity, as they provide the most information about the user.

\section{Unsolved Problem: Security of Noise-Based Authentication}

We have demonstrated an example of unique noise fingerprints from a single user’s fingerprint, face, and eye, drawn from image data. While these noise fingerprints are unique, an adversary could feasibly replicate these noises, potentially compromising the security of traditional image-based biometric systems.

Given the replicability of noise fingerprints from images, we propose thermal noise as an alternative source of biometric data that could offer greater security. Unlike image-based noise, thermal noise is generated intrinsically by the user’s physical properties, such as body heat and blood flow. This raises the following questions: 

\begin{itemize}
\item{Can thermal noise provide a level of robustness and replicability that image-based noise fingerprints lack?}

\item{Are there inherent complexities in thermal noise that make it harder to duplicate compared to image-based noise?}

\item{How stable and consistent is thermal noise under different conditions, e.g. across different temperatures or when a user is physically stressed?}

\item{How might this shift the landscape of biometric security (i.e. what new vulnerabilities could arise, and how might they be mitigated)?}
\end{itemize}

\end{document}